\pdfoutput=1
\documentclass{aastex631}

\usepackage{amsmath}

\begin{document}

\title{The Optical Origin of the Mass-Sheet Transformation\footnote{%
The arXiv version includes minor textual clarifications relative to the published RNAAS version; no results or conclusions are changed.}}

\author[0000-0002-1155-2295]{Marc V. Gorenstein}
\affiliation{Independent Researcher, Natick, MA}

\begin{abstract}

In gravitational lensing, the Mass-Sheet Transformation (MST)—or mass-sheet degeneracy—leaves image positions unchanged while scaling magnifications and time delays. The transformation scales the lens mass distribution and superposes a uniform mass sheet, but this formulation offers no clear physical interpretation. Here I show that the MST follows directly from a scaling symmetry that becomes apparent when the ray–trace relation is written in proper–distance coordinates. In this form, the ray–trace relation isolates a geometric focusing term. Subtracting this term from the deflection law defines the Image-Selection Relation (ISR), which determines image positions, magnifications, and differential time delays. The ISR exhibits a scaling symmetry that leaves image positions unchanged while scaling magnifications and time delays. Restoring the geometric focusing term then gives a ray–trace relation related to the original one by the Mass-Sheet Transformation.

\end{abstract}

\section*{Introduction}

E.~E.~Falco, M.~V.~Gorenstein, and I.~I.~Shapiro (1985) showed that for a gravitational lens confined to a single deflector plane, scaling the mass distribution and superposing a uniform mass sheet leave image positions unchanged while scaling magnifications and time delays. 
This \textit{Mass-Sheet Transformation} (MST) revealed that image observables alone do not fix the mass scale of the lens, and therefore cannot determine the Hubble constant $H_0$ without additional constraints. The Mass-Sheet Degeneracy (MSD) remains a principal limitation in strong-lens modeling and is now recognized as one of a broader family of mass-profile and source-position degeneracies (e.g., Suyu et~al.\ 2010; Schneider \& Sluse 2013; Birrer et~al.\ 2020; Chen et~al.\ 2021; Birrer et~al.\ 2024; Khadka et~al.\ 2024). This Note describes the geometric--optical symmetry that underlies the MSD, the dominant---and simplest---degeneracy present in all strong-lens models.

\section*{Ray-Trace Relation}

Following Falco et al.\ (1985), let the angular-diameter distances between the observer--deflector, observer--source, and deflector--source be $D_d$, $D_s$, and $D_{ds}$. The ray--trace relation connects the physical deflection law $\boldsymbol{\Lambda}(\boldsymbol{\Theta})$ to the unlensed source position $\boldsymbol{\Theta_s}$ and the observed image angles $\boldsymbol{\Theta}$:
\[
\boldsymbol{\Lambda}(\boldsymbol{\Theta})
=
(\boldsymbol{\Theta_s} - \boldsymbol{\Theta})\,\frac{D_s}{D_{ds}}.
\]

A change of variables makes the optical structure explicit.
Substituting $\boldsymbol{\Theta}=\boldsymbol{b}/D_d$ and
$\boldsymbol{\Theta_s}=\boldsymbol{b_s}/D_d$ rewrites the relation in
proper-distance coordinates on the deflector plane, where the rays
intersect the lens. The deflection law becomes $\boldsymbol{\Lambda}(\boldsymbol{b})$,
and the ray--trace relation takes the form
\[
\boldsymbol{\Lambda}(\boldsymbol{b})
=
\frac{\boldsymbol{b_s - b}}{\mathbb{D}},
\tag{1}
\]
with
\[
\mathbb{D} \equiv \frac{D_d D_{ds}}{D_s}.
\]
Written in this form, all geometric dependence is carried by $\mathbb{D}$,
the \emph{distance factor} for the lens system. The image positions are the points $\boldsymbol{b_i}$ that satisfy Equation~(1).

Equation~(1) contains the term $-\boldsymbol{b}/\mathbb{D}$, a
geometric constraint imposed by the ray--trace relation. The same term also has
the form of the deflection law of a simple focusing lens of focal length
$\mathbb{D}$. The presence of this term, hereafter referred to as the
\textit{geometric focusing term}, suggests decomposing the deflection
law into the sum of two idealized optical components: a focusing element that
provides the geometric convergence required for imaging, and an image--selection
element that determines which rays form the images. The next section introduces
the focusing element; the following one defines the image--selection element.

\section*{The Einstein Lens}

Falco et al.\ (1985) modeled the focusing term—what is now termed a mass sheet—as a circular disk with uniform surface-mass density $\sigma_0$ covering the image-formation region. A uniform disk acts as a thin lens with focal length $1/(4\pi\sigma_0)$ and deflection angle $\alpha(b) = -4\pi\sigma_0\,b$. 
With $G=c=1$, mass has dimensions of length and surface-mass density has dimensions of inverse length.

\textbf{Definition.}
An \emph{Einstein Lens} is a transparent circular disk on the deflector plane with focal length equal to the distance factor $\mathbb{D}$ and large enough to cover the image-formation region.  Its surface-mass density is
\[
\sigma_E \equiv \frac{1}{4\pi\mathbb{D}} ,
\]
which yields the deflection law
\[
\boldsymbol{\alpha}_E(\boldsymbol{b}) = -\frac{\boldsymbol{b}}{\mathbb{D}} .
\tag{2}
\]

Equations~(1) and~(2) make explicit the focusing structure of gravitational lensing.
Images lie at those points $\boldsymbol{b}_i$ where the physical deflection law $\boldsymbol{\Lambda}(\boldsymbol{b})$ matches that of an Einstein Lens centered on the projected source position. At such points, rays from the source are brought to a common focus at the observer.

Gorenstein et al.\ (1988, Appendix~C) noted that a uniform disk of this form—here termed an Einstein Lens—when centered on the observer–source line, satisfies the condition for forming Einstein-ring images at all radii within the disk.

\section*{The Image-Selection Lens}

Only some rays from the source reach the observer and form the multiple images. Subtracting the Einstein--Lens deflection from the physical deflection isolates the element that selects the image-forming rays---the \textit{Image-Selection Lens} (ISL):
\[
\boldsymbol{\alpha}_I(\boldsymbol{b}) \equiv \boldsymbol{\Lambda}(\boldsymbol{b}) - \boldsymbol{\alpha}_E(\boldsymbol{b}).
\tag{3}
\]
Equation~(3) expresses the physical deflection law as the sum of two elements,
\[
\boldsymbol{\Lambda}(\boldsymbol{b}) = 
\boldsymbol{\alpha}_I(\boldsymbol{b}) + 
\boldsymbol{\alpha}_E(\boldsymbol{b}),
\tag{4}
\]
and substituting this into Equation~(1) gives
\[
\boldsymbol{\alpha}_I(\boldsymbol{b}) +
\boldsymbol{\alpha}_E(\boldsymbol{b})
=
\frac{\boldsymbol{b_s - b}}{\mathbb{D}},
\tag{5}
\]
which is the ray–trace relation now written in terms of the ISL and Einstein Lens.

Equation~(2) defines the deflection law of the Einstein Lens. This deflection
law, $-\boldsymbol{b}/\mathbb{D}$, is the same as the geometric focusing term
that appears in the ray--trace relation. Subtracting Equation~(2) from
Equation~(5) gives the Image-Selection Relation (ISR):
\[
\boldsymbol{\alpha}_I(\boldsymbol{b}) = \frac{\boldsymbol{b_s}}{\mathbb{D}}.
\tag{6}
\]
The image positions are those values of $\boldsymbol{b_i}$ that satisfy
Equation~(6). At these image positions the ISL acts like a prism that deflects
the diverging rays from the source by the same deflection angle,
$\boldsymbol{b_s}/\mathbb{D}$. Equation~(6) is equivalent to Equation~(1): for a
given source position $\boldsymbol{b_s}$, both equations yield the same set of
image positions $\boldsymbol{b_i}$.

Because the ISR is equivalent to the ray--trace relation, it carries all of the image-forming content of the original system. The ISR determines magnifications through the Jacobian of the mapping between $\boldsymbol{b_s}$ and $\boldsymbol{b}$, and differences in the ISL potential at the image positions give the differential time delays.

The ISR also possesses a scaling freedom. Multiplying Equation~(6) by a factor $\varepsilon$ gives
\[
\varepsilon\,\boldsymbol{\alpha}_I(\boldsymbol{b}) =
\frac{\varepsilon\,\boldsymbol{b_s}}{\mathbb{D}} .
\tag{7}
\]
This scaling modifies both the ISL deflection and the source position, yet any image position $\boldsymbol{b_i}$ that satisfies Equation~(6) also satisfies Equation~(7). Thus the image positions remain invariant under a uniform scaling of the ISR---a symmetry at the heart of gravitational lensing.

Because the source position scales as $\boldsymbol{b_s} \rightarrow \varepsilon \boldsymbol{b_s}$, the total magnification scales as $\varepsilon^{-2}$, and the potential delay---and therefore the differential time delay---scales as $\varepsilon$.

\section*{The Mass-Sheet Transformation}

An optical system that simulates the formation of multiple images can be constructed by combining an Image-Selection Lens (ISL) with an Einstein Lens (EL). Given a set of positions where images are to appear, construct an ISL whose deflection takes the same constant value at those points. The observer--deflector--source distances fix the distance factor and the EL focal length; together with the ISL’s constant deflection they determine the source position and the ISR. Adding the EL focusing terms to the ISR yields the ray--trace relation for this ISL--EL system, whose solutions are the selected image positions.

I apply this general construction to the specific case at hand by adding Equations~(7) and~(2), which yields
\[
\varepsilon\,\boldsymbol{\alpha}_I(\boldsymbol{b}) + \boldsymbol{\alpha}_E(\boldsymbol{b})
=
\frac{\varepsilon\,\boldsymbol{b_s} - \boldsymbol{b}}{\mathbb{D}}.
\tag{8}
\]
Equation~(8) is the ray--trace relation for the system constructed from the scaled ISL and the Einstein Lens.

Substituting Equation~(3) into Equation~(8) rewrites the ray--trace relation in terms of the physical deflection law~$\boldsymbol{\Lambda}(\boldsymbol{b})$, which then becomes
\[
\varepsilon\,\boldsymbol{\Lambda}(\boldsymbol{b})
+
(1-\varepsilon)\,\boldsymbol{\alpha}_E(\boldsymbol{b})
=
\frac{\varepsilon\,\boldsymbol{b_s} - \boldsymbol{b}}{\mathbb{D}}.
\tag{9}
\]
Equation~(9) has the same form as Equation~(1), and one obtains Equation~(9) from Equation~(1) through the substitutions
\[
\boldsymbol{\Lambda}(\boldsymbol{b})
\;\Rightarrow\;
\varepsilon\,\boldsymbol{\Lambda}(\boldsymbol{b}) +
(1-\varepsilon)\,\boldsymbol{\alpha}_E(\boldsymbol{b}),
\qquad
\boldsymbol{b_s}
\;\Rightarrow\;
\varepsilon\,\boldsymbol{b_s}.
\]
These replacements rescale the physical deflection law, add an Einstein--Lens element scaled by $1-\varepsilon$, and scale the source position by $\varepsilon$, leaving the image positions fixed while magnifications scale as $\varepsilon^{-2}$ and differential time delays as $\varepsilon$.

These are the transformations introduced by Falco, Gorenstein, and Shapiro (1985). In their formulation the transformations were introduced and then used to derive the image invariance and scalings; here they emerge directly from the optical properties of the ISL and its scaling symmetry.

\section*{Conclusion}
The observable properties of multiple-image systems constrain both the mass distribution of the deflector and the cosmological distance scale. Yet the ISR’s scaling freedom—responsible for the classic Mass-Sheet Degeneracy associated with the MST—is a geometric--optical symmetry intrinsic to any gravitational lens system. The MSD, together with the broader family of mass-profile and source-position degeneracies, limits the accuracy of $H_0$ inferred from measurements of time delays. Breaking these degeneracies requires external data---stellar kinematics and information on the deflector, its environment, and the line of sight---beyond that provided by the images. This Note identifies the geometric--optical symmetry underlying the MST, the dominant degeneracy that every strong-lens analysis must take into account.

\begin{acknowledgments}
I thank Paul Schechter for his encouragement to publish this note.
\end{acknowledgments}


\begin{thebibliography}{}

\bibitem[Birrer et al.(2020)]{Birrer2020}
Birrer, S., Shajib, A.~J., Galan, A., et al.\ 2020, \aap, 643, A165

\bibitem[Birrer et al.(2024)]{Birrer2024}
Birrer, S., Millon, M., Sluse, D., et al.\ 2024, \ssr, 220, 48

\bibitem[Chen et al.(2021)]{Chen2021}
Chen, G.~C.-F., Fassnacht, C.~D., Suyu, S.~H., et al.\ 2021, \aap, 652, A7

\bibitem[Falco, Gorenstein, \& Shapiro(1985)]{FGS1985}
Falco, E. E., Gorenstein, M. V., \& Shapiro, I. I. 1985, ApJL, 289, L1

\bibitem[Gorenstein et al.(1988)]{Gorenstein1988}
Gorenstein, M.~V., Falco, E.~E., \& Shapiro, I.~I.\ 1988, \apj, 327, 693

\bibitem[Khadka et al.(2024)]{Khadka2024}
Khadka, N., Birrer, S., Leauthaud, A., \& Nix, H.\ 2024, \mnras, 533, 795

\bibitem[Schneider \& Sluse(2013)]{SchneiderSluse2013}
Schneider, P., \& Sluse, D.\ 2013, \aap, 559, A37

\bibitem[Suyu et al.(2010)]{Suyu2010}
Suyu, S.~H., Marshall, P.~J., Auger, M.~W., et al.\ 2010, \apj, 711, 201



\end{thebibliography}
\end{document}